\documentclass[prl,graphicx,amssymb]{revtex4}
  \usepackage{color}
\usepackage{graphicx}
\begin{document}

\title{Comment on ``One-state vector formalism for the evolution of a quantum state through
nested Mach-Zehnder interferometers''}
\author{ L. Vaidman}
\affiliation{ Raymond and Beverly Sackler School of Physics and Astronomy\\
 Tel-Aviv University, Tel-Aviv 69978, Israel}
\begin{abstract}
Bartkiewicz {\it et al.} [Phys. Rev. A {\bf 91}, 012103 (2015)] provided an alternative analysis of  experiment
performed by Danan {\it et al.} [Phys. Rev. Lett. {\bf 111}, 240402 (2013)] which presented surprising   evidence regarding the past of photons passing through an interferometer. They argued that   the quantity used by Danan {\it et al.}  is not a suitable which-path witness, and proposed  an alternative. It is argued that the quantum and classical analyses of Bartkiewicz {\it et al.} are inconsistent and both are inappropriate for describing the past of photons in a properly working interferometer.
\end{abstract}

\maketitle

The experiment \cite{Danan} continues to be in the center of a controversy. Bartkiewicz {\it et al.} \cite{Bart} (BCJLSS) argued that they ``shed more light on the ongoing discussion regarding the legitimacy of the experimental data and their interpretation.'' They write that ``theoretical
calculations and an interpretation of the experimental results
using only the standard one-state vector quantum-mechanical approach are still lacking,'' and that they ``establish a more reliable which-path witness and show that it yields well-expected outcomes of the experiment.''

 Danan {\it et al.} \cite{Danan} ``asked'' photons where have they been inside an interferometer. BCJLSS considered the same interferometer, but proposed to ask the photons in a different way. In fact, in their paper they have two  proposals: one quantum and one classical. BCJLSS claim that their quantum and classical analyses predict the same results reproducing the experimental data     presented in   Fig. 3a and 3b of \cite{Danan}  (BCJLSS Fig. 1a and 1b), but differ in the experiment presented in Fig. 3c. (their Fig. 1c). I will denote these experiments as  cases (a), (b) , and (c).
I find their analysis incorrect in many ways.

For describing  interaction of photons with vibrating mirrors BCJLSS introduce ``frequency modes'' which in principle allow  to know which mirror the photon interacted with. Their quantum proposal is that  instead of placing the quad-cell detector at the  output of the interferometer, the light emitted during the run (the duration of which was 1 sec) is stored and then a quantum measurement of the mode is performed. The measurement should provide one of the existing modes which specifies the mirror a particular photon interacted with.

This method is  conceptually  incorrect for analysing position of particles inside an interferometer. Danan {\it et al.} wanted to know where the photons were  in an interference experiment. A measurement telling us an exact  path of the photon invariably destroys interference, so it cannot tell us much about photons passing through properly working interferometer. This is why Danan {\it et al.} used weak measurements in which the information comes from many photons. It provided almost no  information about location of any individual photon.

But how can this difference arise when the setups are identical until the final measurements performed at the output port, after the interference took place? When the photons pass through the interferometer, the information is written in their transversal degree of freedom. How does it happen that the way this information is observed makes so big difference?   BCJLSS do not mention that most of the photons in the output port will not be detected in one of the frequency modes they mentioned, but in the zero frequency mode corresponding to null information about location of photons inside the interferometer. The amplitude of vibration in the experiment is very tiny, so in the output port almost all photons will have zero frequency mode. The measurements of BCJLSS provide no direct information about where these photons were.

A possible approach is to consider the full ensemble and to argue that  fractions of photons with observed frequency modes tell us where the photons with observed zero frequency mode were present.  Danan {\it et al.} also made an assumption of a similar type: the information obtained from the ensemble was interpreted as relevant for every single photon. BCJLSS could argue that the photons with detected mirror frequencies were disturbed by their measurements, but they allowed to know where  were the zero mode photons, which were the majority in this interference experiment.

Note, however, that BCJLSS are  vulnerable to the following line of criticism: The photons, disturbed by the measurement, cannot provide a reliable information about the undisturbed photons, since disturbance destroys interference. This criticism is less effective for the experiment  \cite{Danan} in which the information is taken from all photons, while the measurement disturbance spoils the interference of every photon only very little.

Curiously, the success of BCJLSS method has an explanation in the two-state vector formalism \cite{AV90}. It asserts that the photon was present in the overlap of the  forward and backward evolving wave functions. A mirror can create a mode with its frequency if the forward evolving wave reaches it. If the backward evolving wave reaches the mirror, it means that the photon mode created there can reach the detector.

Besides the conceptual disadvantage of BCJLSS method, I argue that they do not analyze it correctly. Note, that case (a) of \cite{Danan} and case (a) of BCJLSS are different by a relative phase of $\pi$ of path $C$, such that the latter does not correspond to a full constructive interference of the interferometer as it was  in the experiment \cite{Danan}. However, in both cases the analysis  of \cite{Danan} makes the same predictions regarding   positions of pre- and post-selected photons inside the interferometer, so we can follow the setup of BCJLSS.

In the experiment, the correct transformation of the photon due to interaction with a vibrating mirror, say mirror $C$, is not $  \hat c_{00000} \rightarrow    \hat c_{00100} $ as it appears in Eq. (2) of BCJLSS, but
\begin{equation}\label{2}
 \hat c_{00000} \rightarrow  \frac{1}{\sqrt{1+\epsilon^2}}(\hat c_{00000} + \epsilon ~ \hat c_{00100}),
\end{equation}
where $\epsilon \ll 1$. Then, the correct output state in the Fock space (instead of Eq. (9) of BCJLSS) is:
\begin{equation}\label{9}
  |\Psi_{out} \rangle = {\cal N}  \{  e^{i\varphi}  |1\rangle_{00000} + \epsilon [|1\rangle_{00100} +
  ( e^{i\varphi}-1) (|1\rangle_{00010}+|1\rangle_{00001})   + e^{i\varphi}|1\rangle_{10000}- |1\rangle_{01000}] \} + h.o.t.
\end{equation}
where  the normalization factor ${\cal N}=\frac{1}{3}+O(\epsilon^2)$  corresponds to a single photon in all output ports, and the high order terms  are:
\begin{equation}\label{9hot}
   h.o.t. = {\cal N}  \{ \epsilon^2 [ e^{i\varphi} ( |1\rangle_{10010} + |1\rangle_{10001}) -(|1\rangle_{01010} + |1\rangle_{01001})]
 +\epsilon^3 (e^{i\varphi}|1\rangle_{10011} -|1\rangle_{01011}) \}.
\end{equation}

  The probability of post-selection  in the output port on a particular frequency mode, say mode $A$, is proportional to the weight of the output state projected on the space which has this mode:
 \begin{equation}\label{probA}
 {\rm prob} (A)= |{\rm \bf P_A}|\Psi_{out} \rangle |^2,
 \end{equation}
 where
 \begin{equation}\label{projA}
  {\rm \bf P_A}\equiv \sum_{i,j,k,l=0,1} |1\rangle_{1ijkl}~\langle 1|_{1ijkl} .
 \end{equation}
 This method provides  the results of cases (a) and (b). For case (a),  $\varphi = \pi$,
 \begin{equation}\label{caseA}
 {\rm prob} (A)= {\rm prob} (B)= {\rm prob} (C)=\frac{\epsilon^2}{9}, ~~~~{\rm prob} (E)= {\rm prob} (F) =\frac{4\epsilon^2}{9}.
 \end{equation}
The probability for finding the zero mode  is $\frac{1}{9}$.

 For case (b), corresponding to phase $\varphi = 0$,
\begin{equation}\label{probB}
 {\rm prob} (A)= {\rm prob} (B)= {\rm prob} (C)=\frac{\epsilon^2}{9}, ~~~~{\rm prob} (E)= {\rm prob} (F) =0,
 \end{equation}
while the probability for finding the zero mode in this case is $\frac{1}{9}$, as in case (a).

 The state in the output port (\ref{9}) is very different from the output state obtained by BCJLSS, as seen in their Eq. (9):
 \begin{equation}\label{9orig}
  |\psi_{out} \rangle = \frac{1}{3}~ (   |1\rangle_{00100} - |1\rangle_{01011} + e^{i\varphi} |1\rangle_{10011}).
  \end{equation}
In order to  obtain the  results of the experiment  from their state, BCJLSS introduced a procedure ``formally equivalent to the projection of the output state'' in which they defined the post-selected state in their Eq. (10):
 \begin{equation}\label{projAorig}
  |\Pi_A\rangle= \sum_{i,j,k,l=0,1} |1\rangle_{1ijkl},
 \end{equation}
 and then calculated the square of the value of the scalar product between (\ref{9orig}) and (\ref{projAorig}), their Eq. (11):
 \begin{equation}\label{probAorig}
|\langle{ \Pi_A}|\psi_{out} \rangle |^2.
 \end{equation}

  BCJLSS used  (\ref{9orig}) instead of (\ref{9}),  (\ref{projAorig})  instead of (\ref{projA}), (\ref{probAorig}) instead of (\ref{probA}), and obtained results which are proportional to what I obtained, but I cannot see a justification for their procedure.  Consider, for example, the absence of modes $E$ and $F$ in case (b) when $\varphi =0$. When the output state (\ref{9orig}) has terms $|1\rangle_{10011},  |1\rangle_{01011}$, the   modes $E$ and $F$ must appear  independently on the phase, since these are two orthogonal modes and thus they cannot interfere.
BCJLSS obtained cancelation of the modes performing   scalar product with the post-selected state
 \begin{equation}\label{projBorig}
  |\Pi_B\rangle= \sum_{i,j,k,l=0,1} |1\rangle_{i1jkl},
 \end{equation}
  which has   particular phases between terms with different modes. What is the physical mechanism  which  fixes these phases? (In my analysis, the modes $E$ and $F$ are not seen in case (b) because they appear only in the high order in $ \epsilon$ in the output state, see (\ref{9hot}).)

Another argument against  BCJLSS approach, which is not related to the method of observing output photons, is that in the experiment \cite{Danan}  no  significant disturbance of visibility of the inner interferometer, due to vibration of the mirrors, was observed. The orthogonality of modes $\hat c_{10011}$ and $\hat c_{01011}$ appearing in (\ref{9orig}) is in contradiction with this fact.

Let us turn to the case (c). BCJLSS wrote that their ``theoretical prediction does not match
the experimental data from Ref. \cite{Danan}''.  Their Fig. 1c   tells us  that in this case the photon was in $A$ and/or in $B$. If their method is a reliable which-path witness, as they claim, then Fig. 1c also tells us that the photon was not present in $C$,  $E$ and $F$. Indeed, the BCJLSS measurement method of detecting the frequency modes   provides the following probabilities: \begin{equation}\label{probC}
 {\rm prob} (A)= {\rm prob} (B)=\frac{\epsilon^2}{9}, ~~~~ {\rm prob} (C)={\rm prob} (E)= {\rm prob} (F) =0.
 \end{equation}
But this is not a ``well-expected'' behaviour. It means that the photon did not exist for some periods of time before and  after being in the nested interferometer.
I do not accept this. According to my approach \cite{past},   the photon was where it left a weak trace. The gedanken experiment of case (c), performed on a pre- and post-selected ensemble with finite amplitude of  vibration of mirrors,  the post-selection of, say,  mode $A$ by BCJLSS method, and weak measurement of the trace with external devices, will  show a continuous trajectory which   includes $E$, $A$ and $F$.

Danan {\it et al.}  do show ``unexpected behaviour'' when photon cannot be described by  continues trajectories. This is the case  (b) in which the weak trace of the photon shows continuous trajectory through $C$ together with a separate trace inside the inner interferometer including $A$, and  $B$.

Assuming ideal devices, the probability for finding the zero mode in  case (c) is 0. So, in such an experiment detector $D$ observes only disturbed photons. There are no photons performing interference experiment about which we can make statements regarding their position.

Note that in \cite{Danan}, the case (c) was introduced not as an experiment which is supposed to tell where a pre- and post-selected photon was, but as a test that the experimental system, especially for the case (b), worked properly. In the limit of zero disturbance,  no pre- and post-selected photons are  expected, and indeed, there was no signal beyond the noise of the system, see more discussion in \cite{Sal,ReplySal}.

BCJLSS also suggest another, classical way of asking photons where they have been. No need for storing photons one light second long and performing complicated mode measurements. Moreover, even the quad-cell detector is not needed, just a single detector measuring intensity as a function of time $I_T(t)$, and a computer which makes Fourier analysis. BCJLSS argue that the Fourier transform of the total intensity ``is a more reliable which-path witness and it corresponds to their quantum model''. Note, however, that this is {\it my} reading of the BCJLSS classical proposal, based on the words in the text ``spectrum of the overall intensity''. One of their equations defines the term  $I_T(f)$, which apparently is not the Fourier transform of $I_T(t)$,  and for which there is no proposal for actual experiment.

BCJLSS admitted that they  ``did not keep track of the normalization factors''.  Intensity modulation is tiny, it is of the order $\epsilon^2$, where $\epsilon$ is the ratio between the shift of the beam due to vibration of a mirror and the width of the beam. This can be learned from the leakage of nested interferometer in case (b), see Eq. (8) of \cite{past}. This puts a serious question mark on the feasibility of this method even in its simple version. However, I argue that the total intensity does not provide a which-path witness corresponding to the quantum model even with gedanken devices with unlimited precision.

The simplest way to see that BCJLSS new which-path witness is not reliable, is that in a simple case when instead of an interferometer there is only one mirror on the path of the photon, the vibration  of the mirror does not cause  a change in the total intensity, so it does not tell us that the photon bounced off the mirror.

 I will show now that   BCJLSS  proof, that for the nested interferometer their classical which-path witness provides the same predictions as the quantum method,  arises from a mathematical error.
The starting point of the BCJLSS calculation of the intensity modulation is their first (unnumbered) equation of Section III:
\begin{equation}\label{psiY}
\Psi(y,t) \propto \kappa e^{-(y-d_C)^2} -e^{-(y-d_A-d_E-d_F)^2}+e^{i\varphi}e^{-(y-d_B-d_E-d_F)^2},
 \end{equation}
where $\kappa = 1$ for cases (a) and (b), and $\kappa = 0$ for case (c). Then, the which-path witness is given by the Fourier transform of the total intensity, $I_T(t)= \int_{-\infty}^{\infty} |\Psi(y,t)|^2dy$.
To obtain their results, BCJLSS made approximations of (\ref{psiY}) for various cases.

For case (c),  the approximation is $2ye^{-y^2}(d_A-d_B)$ (their Eq. (18)). The time modulations of $d_A$ and $d_B$ lead to equal contributions to  $I_T(f)$ at the frequencies of mirrors $A$ and $B$ corresponding to their Fig. 1c. (BCJLSS disregarded doubling the frequencies appearing in the spectrum of the total intensity relative to  the frequencies of the time modulations  of the beam shifts. Since one characterises the other, I, as BCJLSS, will characterize the signal by the frequencies of the original time modulations of the beam positions.)  For cases (a) and (b), the BCJLSS approximations, given by their Eqs. (15) and    (17), also allowed to reproduce the results of their quantum model presented in their Fig. 1a and 1b, which I reproduce in Fig. 1A. However, exact calculations based on (\ref{psiY}) show different results, see Fig. 1B.

\begin{figure}[t]
 \begin{center} \includegraphics[width=9.5cm]{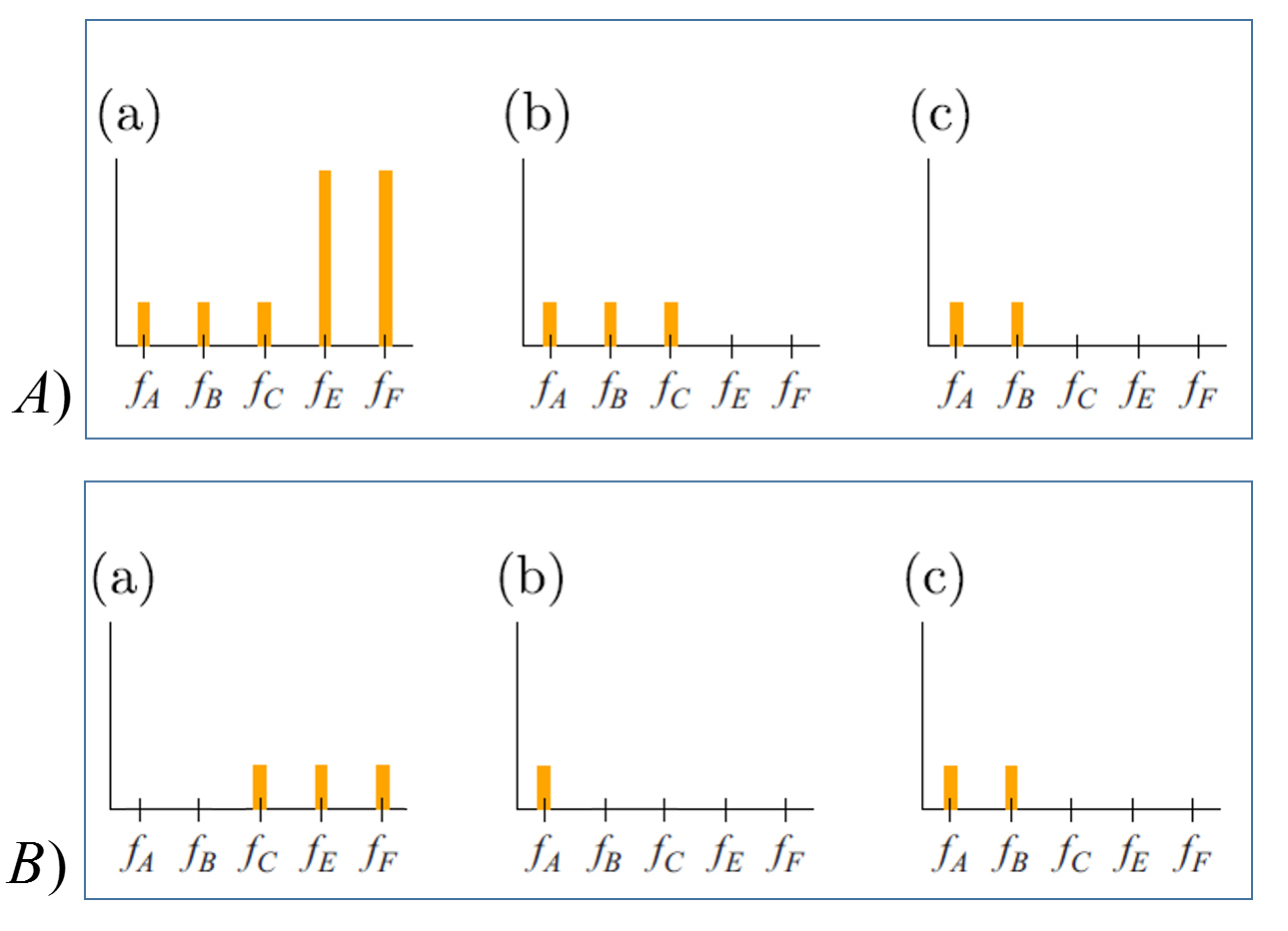}\\
      \caption{(A) Power spectrum of the signal for different cases predicted by quantum and classical models as described by BCJLSS. (B) Power spectrum of the BCJLSS which-path witness  for different cases calculated on the basis of the exact expression (\ref{psiY}). } \label{2}\end{center}
\end{figure}

Indeed, in case (b), beam $A$ essentially cancels beam $B$, so beam $C$ has nothing to interfere with and, therefore, its shift does not change the total intensity. The same for beam $B$, since beam $C$ cancels beam $A$. Beam $A$ interferes   with the combined beam of $B$ and $C$ and, therefore, its position does change the total intensity of the output port. Thus, the only frequency which appears is $f_A$. (Mirrors $E$ and $F$   shift   the beam from the nested interferometer   whose intensity  at the detector is almost zero, so their effect is negligible in this case.) Similarly, in case (a), frequencies $f_A$ and $f_B$  do not appear, since beams from $A$ and $B$ have nothing to interfere with. Mirrors $E$ and $F$  now shift the beam which is twice as strong as beam $C$, but since for total intensity it does not matter if the small beam shifts relative to a large beam or vice versa, the power of the signals at $f_E$  and $f_F$ is the same as at $f_C$.

It is a curiosity that the approximation in case (c) (their Eq. (18)) misled  BCJLSS regarding null signal   $\Delta I(t)$, the output of the quad-cell detector, which motivated them to look for another which-path witness. $\Delta I(t)$ is identically zero when calculated based on the approximation, but non zero (albeit very small) when calculated based on the exact expression (\ref{psiY}). In the Danan {\it et al.} experiment, the number of detected photons in case (c) was too small to see the signal given the background noise. The  analysis of  $\Delta I(t)$ in the ideal experiment does provide information that detected photons in case (c) were in $A$ and in $B$, but it also  tells us that they were in $E$ and in $F$.

The past of a photon in an interferometer is a subtle issue both theoretically and experimentally. Quantum mechanics does not have unambiguous definition for a position  of a pre- and post-selected particle \cite{past,morepast}. The experiment \cite{Danan} does show correctly  where the photons were, when the  definition of a location  of a pre- and post-selected particle is all places where it leaves a trace.   It is not a direct measurement of the trace left by photons on other systems. It uses a degree of freedom of the photon itself which makes the experiment much easier to perform,  but more difficult to justify. Some variations of this experiment, like different analysis of the output photons  proposed by  BCJLSS, or apparently harmless insertion  of a Dove prism   \cite{Jordan},  might not properly record and/or read this trace.

In light of the above, I argue that BCJLSS analysis does not help to understand the past of photons passing through interferometers. Their which-path witness provides new predictions only in case (c) in which there are no undisturbed photons passing through the interferometer. I also do not think that calculations  of the experimental results of Danan {\it et al.} using only the standard one-state vector quantum-mechanical approach are  lacking. The original Letter \cite{Danan} with its supplemental material and  preceding theoretical article \cite{past} provide a correct  one-state vector quantum description. It was repeated in  details by Saldanha \cite{Salda}, and recently repeated again even with more details by Poto\v{c}ek  and Ferenczi \cite{Poto}. However, the analysis of the physical meaning of the results and the consensus regarding their interpretation are still wanted.

This work has been supported in part by the Israel Science Foundation  Grant No. 1311/14  and the German-Israeli Foundation  Grant No. I-1275-303.14.

\end{document}